\documentclass[published]{epl}

\renewcommand{\onefigure}[1]{%
  \stepcounter{epl@cnt@fig}%
  \hbox to\hsize{\hfill\includegraphics[width=0.55\textwidth]{#1}\hfill}%
}

\begin{document}


\title{Non-Kramers freezing and unfreezing of tunneling in the biaxial
  spin model}
\shorttitle{Unfreezing of tunneling in spin models}

\author{
  E. M. Chudnovsky\inst{1} \thanks{E-mail:
    \email{chudnov@lehman.cuny.edu}}
  \and X. Mart\'\i nez-Hidalgo\inst{1,2} \thanks{E-mail:
    \email{xavim@ubxlab.com}}
}
\shortauthor{E. M. Chudnovsky \etal}

\institute{
  \inst{1} Department of Physics and Astronomy - City University of New
  York--Lehman College, Bedford Park Boulevard West, Bronx, New
  York 10468-1589, USA \\
  \inst{2} Departament de F\'\i sica Fonamental - Universitat de
  Barcelona, Diagonal 647, 08028 Barcelona, Spain
}



\issue{50}{3}{2000}{395}{1 May 2000}
\rec{2 December 1999}{15 February 2000}

\maketitle

\begin{abstract}
  The ground state tunnel splitting for the biaxial spin model in the
  magnetic field, ${\cal H}=-D S_{x}^{2} + E S_{z}^{2} -
  g\mu_{B}S_{z}H_{z}$, has been investigated using an instanton
  approach. We find a new type of spin instanton and a new quantum
  interference phenomenon associated with it: at a certain field,
  $H_{2}=2SE^{1/2}(D+E)^{1/2}/(g{\mu}_{B})$, the dependence of the
  tunneling splitting on the field switches from oscillations to a
  monotonic growth. The predictions of the theory can be tested in
  Fe$_{8}$ molecular nanomagnets.
\end{abstract}


During the last years much effort, both theoretical and experimental,
has been devoted to the study of macroscopic quantum tunneling (MQT)
in spin systems\cite{qtm,book}.
Crystals formed by weakly interacting identical magnetic molecules,
like Mn$_{12}$-acetate and an octanuclear iron cluster Fe$_{8}$, have
been the objects of most intensive recent research.
EPR \cite{Barra} and neutron scattering \cite{Caciuffo} data show that
an Fe$_{8}$ cluster can be described by the biaxial spin Hamiltonian,
\begin{equation}
  \label{hamilton}
  {\cal H}=-D S_{x}^{2} + E S_{z}^{2} - g\mu_{B}\vect{S}\cdot\vect{H}\;,
\end{equation}
where $S=10$, $D\approx 0.23$--$0.27\un{K}$, and $E\approx
0.093\un{K}$ \cite{Barra}.
Macroscopic magnetic measurements \cite{Sangregorio} have revealed
resonant spin tunneling in Fe$_{8}$, confirming the above model and
values of the constants.
At $H=0$ this model has been studied by a number of
authors\cite{ensc86,hesu86,chgu88,gara91}.
The tunnel splitting ${\Delta}$ of the ground state in the case $H\neq
0$ has been theoretically studied using the instanton method by Garg
\cite{Garg}, who discovered an interesting topological effect:
oscillation of ${\Delta}$ on the magnetic field applied along the hard
magnetization axis. Such oscillations have been recently observed in
Fe$_8$ by Wernsdorfer and Sessoli \cite{wese99}. The origin of this
effect is in the quantum interference of different instanton paths,
suggested earlier, within the context of MQT, by Loss, DiVincenzo, and
Grinstein \cite{LDG} and by von Delft and Henley\cite{VH}. 
This oscillatory behaviour has also been derived using standard
perturbation theory by Weigert\cite{weig94}.
The model of Garg has elucidated the fact that the freezing of
tunneling need not be related to Kramers's degeneracy.  To illustrate
his point, Garg studied the field range $0<h<h_{2}=(1-\lambda)^{1/2}$,
where $h=H/H_{c}$, $H_{c}=2S(D+E)/(g\mu_{B})$ is the critical field at
which the energy barrier disappears, and ${\lambda}=D/(D+E)$. In that
field range the tunneling is quenched whenever \cite{Garg}
\begin{equation}
  \label{frzfields}
  h = h_{2}(S-n-1/2)/S\;, \quad n=0,1,\ldots,2S-1.
\end{equation}
This was discovered before the relevant system, Fe$_{8}$, was known.
Meantime, for Fe$_{8}$ ${\lambda}\approx$~0.71--0.75 and, thus, the
above field range covers only the lower half of the range,
$0<H<H_{c}\approx 4.8$--$5.5\un{T}$, available for the tunneling
studies. In this Letter we have considered the remaining field range,
$h_{2}<h<1$, and found another interesting topological effect:
switching from oscillations to the monotonic growth of the tunnel
splitting on the field. We then give explanation to such a behavior
within a continuous spin model containing the Wess-Zumino-Berry
term\cite{Fradkin}.
Hamiltonian (\ref{hamilton}) has been recently studied by Kou et
al.\cite{kou99} who found exact instantons at $h<h_2$ but did not
attempt to solve the problem in the field range $h_{2}<h<1$. We show
that there exists a new type of spin instanton in that range, that
involves motion in both imaginary and real time, which is ultimately
responsible for the above peculiar dependence of the tunnel splitting
on the field.
Complex-time instantons have been used for some time to study barrier
penetration effects using the path integral
formalism\cite{mcla72,patr81,weha82}.
To our knowledge, though, this is the first time that such a
mixed-time trajectory is needed in the study of tunneling in spin
systems.
Garg's and our predictions can be quantitatively tested,
without any fitting parameters, in experiments on Fe$_{8}$.

\begin{figure}
  \onefigure{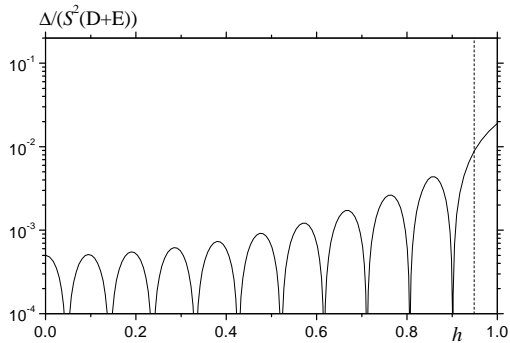}
  \caption{The tunnel splitting $\Delta$ as a function of field
    $h$ at $S=10$ and $\lambda=0.1$. Dashed line marks the unfreezing
    field $h_{2}=0.949$.}
  \label{frzfig1}
\end{figure}

\begin{figure}
  \onefigure{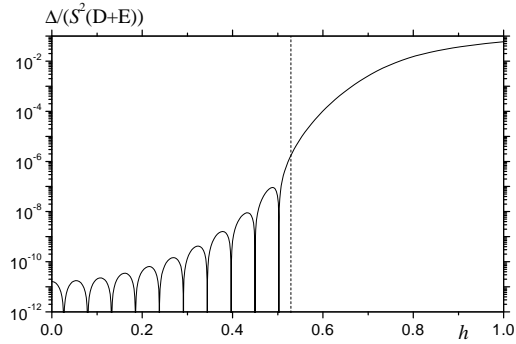}
  \caption{The tunnel splitting $\Delta$ as a function of field
    $h$ for Fe$_{8}$, $S=10$ and $\lambda=0.72$. Dashed line marks the
    unfreezing field $h_{2}=0.529$.}
  \label{frzfig2}
\end{figure}

\begin{figure}
  \onefigure{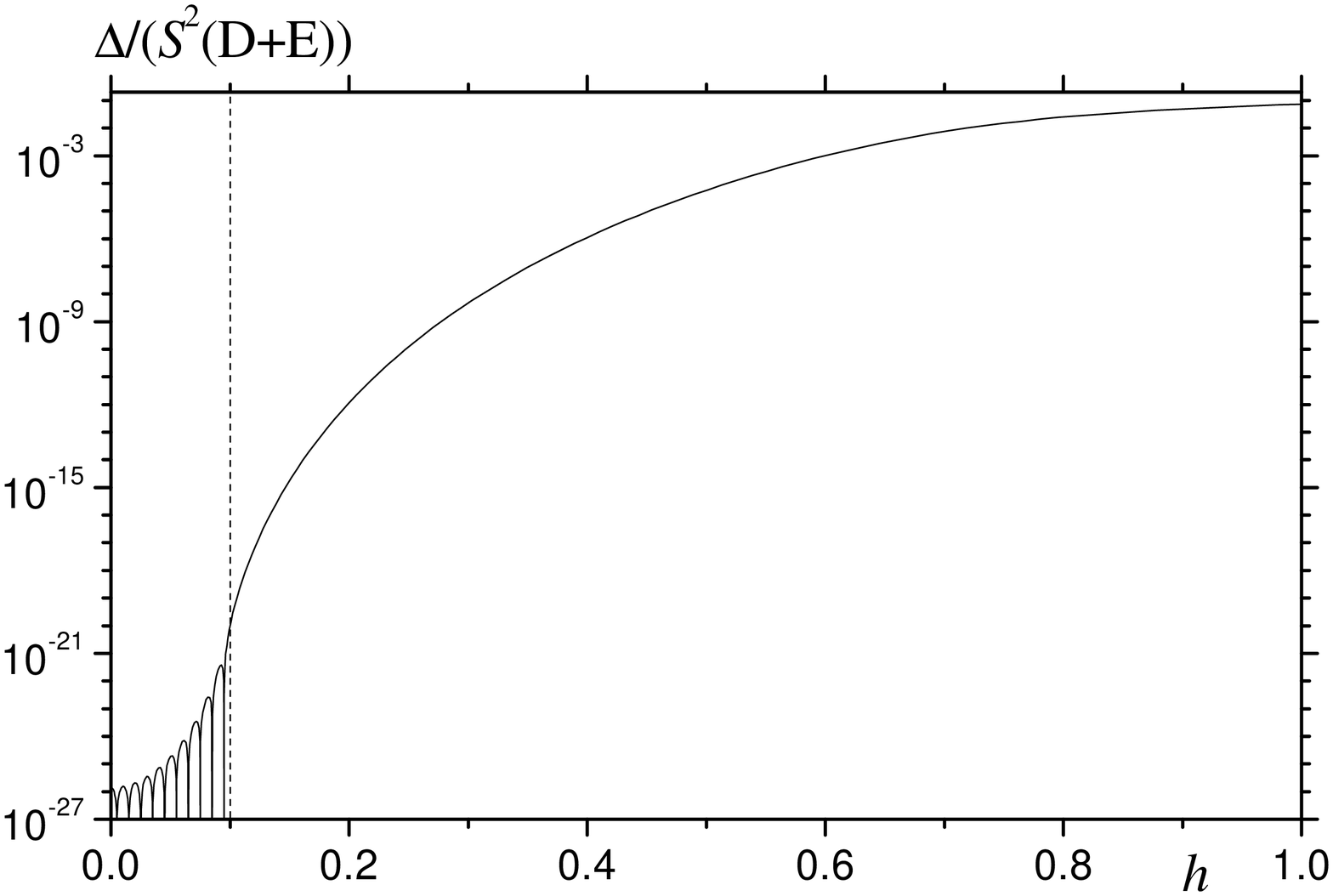}
  \caption{The tunnel splitting $\Delta$ as a function of field
    $h$ at $S=10$ and $\lambda=0.99$. Dashed line marks the unfreezing
    field $h_{2}=0.1$.}
  \label{frzfig3}
\end{figure}

Before studying instantons we will demonstrate the reality of the
effect by performing the numerical diagonalization of the Hamiltonian
(\ref{hamilton}) for $S=10$ in the field applied along the hard axis
$Z$ (see also ref.\ \cite{delb99}).
Figures~\ref{frzfig1}--\ref{frzfig3} show the results of numerical
computations of the ground state tunnel splitting as a function of $h$
for three different values of $\lambda$: small $\lambda$, intermediate
$\lambda$ that corresponds to Fe$_{8}$, and large $\lambda$.  It can
be clearly seen from the figures that Garg's oscillations exist below
a certain field $H_{2}$.  Above that field there are no oscillations;
the tunnel splitting grows linearly with the field. The value of
$H_{2}$ increases with decreasing $\lambda$.  Below, we obtain this
effect via instanton method and show that the crossover field is given
by $h_{2}(\lambda)=(1-\lambda)^{1/2}$.

In the continuous approach the tunnel splitting can be computed as the
functional integral
\begin{equation}
   \label{pathint}
   \oint{\cal D}[\cos\theta(t)]{\cal D}[\phi(t)]
   \exp\left(\frac{i}{\hbar}\int\drm{}t{\cal
       L}[\theta(t),\phi(t)]\right)
\end{equation}
over closed trajectories which describe motion of a fixed-length
vector of the magnetic moment $\vect{M}$ in spherical coordinates
$(\theta(t),\phi(t))$. Here ${\cal L}$ is the Lagrangian of the
magnetic system,
\begin{equation}
  \label{lagrang}
  {\cal L}=\frac{M}{\gamma}(\cos(\theta) - 1)\dot{\phi} -
  {\cal H}(\theta,\phi)\;,
\end{equation}
and ${\cal H}$ is the energy of the system, which includes the magnetic
anisotropy and the Zeeman term,
\begin{eqnarray}
  \label{energy}
  {\cal H}(\theta,\phi) &=&
  -\frac{1}{2}Mk_{\parallel}{\sin}^{2}(\theta){\cos}^{2}(\phi) +
  \frac{1}{2}Mk_{\perp}{\cos}^{2}(\theta) \nonumber \\
  && - MH\cos(\theta) + \frac{1}{2}M\left(k_{\parallel} +
    \frac{H^{2}}{k_{\perp} + k_{\parallel}}\right)\;.
\end{eqnarray}
Here $\gamma=g\mu_{B}/{\hbar}$ is the gyromagnetic ratio, $H$ is the
external field, and $k_{\perp}$ and $k_{\parallel}$ are the hard and
easy anisotropy constants, correspondingly.  The correspondence with
the parameters of the Hamiltonian (\ref{hamilton}) is the following
\[
k_{\parallel}=\frac{2SD}{g\mu_{B}}\;,\quad
k_{\perp}=\frac{2SE}{g\mu_{B}}\;,\quad
H_{c}=k_{\perp} + k_{\parallel}\;,
\]
\begin{equation}
  \vect{S}=\frac{\vect{M}}{\hbar\gamma}\;,\quad
  \lambda=\frac{k_{\parallel}}{k_{\parallel}+k_{\perp}}\;.
\end{equation}
For $H<H_{c}$, the magnetic energy ${\cal H}(\theta,\phi)$ has two equivalent
minima at $\phi=0,\pi$ with $\cos(\theta)=H/H_{c}$. A constant has
been added to make the energy of the minima zero.
The following observation is relevant to the calculation provided
below. The energy (\ref{energy}), besides having the minima mentioned
above, also has two equivalent extrema at ${\phi}={\pm}{\pi}/2$ and
$\cos{\theta}=h/h_{1}$, where $h_{1}=1-\lambda$. At $0<\,h<\,h_{1}$
these are the maxima of the energy.  They would become local minima of
the energy at $h_{1}<\,h<\,h_{2}$ and global minima at $h_{2}<\,h<\,1$
if one allowed for complex ${\theta}$, because
$\cos{\theta}=h/h_{1}>1$ in that field range. We shall see that this
is the case for the effective potential in the quantum problem.

\begin{figure}
  \onefigure{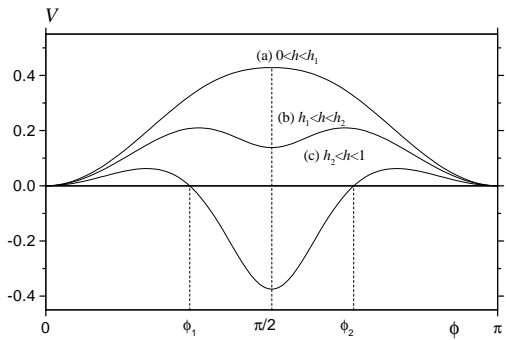}
  \caption{The potential $V(\phi)$ at $\lambda=0.72$ ($h_{1}=0.28$ and
    $h_{2}=0.529$) for three different ranges of the field. The chosen
    fields are: (a) $h=0.2$, (b) $h=0.45$, and (c) $h=0.7$.}
  \label{frzfig4}
\end{figure}

After Gaussian integration over $\cos(\theta)$ in eq.\ 
(\ref{pathint}), the remaining functional integral over $\phi$
contains the imaginary-time effective action
\begin{eqnarray}
  \label{effaction}
  \frac{I}{\hbar} &=& \frac{1}{\hbar} {\int}d\tau {\cal L}_{E} =
  -iS{\int}d\phi \left[ A(\phi)-1 \right] \nonumber \\
  && + S\lambda^{1/2}{\int}d
  \tau' \left[\frac{1}{2} M(\phi) \dot{\phi}^{2}_{\tau'} +V(\phi)
  \right]\;,
\end{eqnarray}
where ${\cal L}_{E}=-{\cal L}(t\rightarrow -i\tau)$ is the Euclidean
Lagrangian, and $\tau'=\tau\gamma [k_{\parallel} (k_{\perp} +
k_{\parallel})]^{1/2}$ is the dimensionless imaginary time. The
functions $A(\phi)$, $M(\phi)$ and $V(\phi)$ are given by
  \begin{equation}
    A(\phi) = \frac{h}{1-\lambda\sin^{2}(\phi)}\;,
  \end{equation}
  \begin{equation}
    M(\phi) = \frac{1}{1-\lambda\sin^{2}(\phi)}\;,
  \end{equation}
and
  \begin{equation}
    V(\phi) = \frac{1}{2}\sin^{2}(\phi) \left[1 -
      \frac{h^{2}}{1-\lambda\sin^{2}(\phi)} \right]\;.
  \end{equation}
The action (\ref{effaction}) is equivalent to that describing the
motion of a particle of mass $M(\phi)$ in an inverted scalar potential
$-V(\phi)$ and a ``vector'' potential $A(\phi)-1$.  The shape of the
potential $V(\phi)$ for three different ranges of the field is shown
in fig.~\ref{frzfig4}. For $h<h_{1}=1-\lambda$, the potential looks
like a regular barrier between $\phi=0$ and $\phi=\pi$, with a maximum
at $\phi=\pi/2$. For $h_{1}<h<h_{2}=(1-\lambda)^{1/2}$, the maximum at
$\phi=\pi/2$ becomes a local minimum, though still higher than the
global minima at $\phi=0,\pi$. For $h_{2}<h<1$, the minimum at
$\phi=\pi/2$ becomes the global minimum.

For $0<h<h_{2}$ one can find the imaginary-time instanton trajectories
for the particle moving from $\phi=0$ to $\phi=\pm\pi$ in the inverted
potential. For these trajectories, the first integral in eq.\ 
(\ref{effaction}) gives an imaginary Wess-Zumino-Berry contribution to
the action,
\begin{equation}
  \label{topol}
  \frac{i}{\hbar}S I_{WZB}^{\pm} = -iS{\int}\drm{}\phi \left[ A(\phi)-1
  \right] = \pm i\pi S \left(1 - \frac{h}{h_{2}}\right)\;.
\end{equation}
The interference of these two trajectories in the functional integral
gives a factor
\begin{equation}
  e^{\frac{i}{\hbar}I_{WZB}^{+}} + e^{\frac{i}{\hbar}I_{WZB}^{-}}=2
  \cos\left[\pi S \left(1 - \frac{h}{h_{2}}\right)\right]\;,
\end{equation}
which is responsible for the non-Kramers freezing of tunneling at
fields satisfying eq.\ (\ref{frzfields})\cite{Garg}.

However, at $h_{2}<h<1$, one cannot find a trajectory in
imaginary-time connecting $0$ and $\pi$. There seems to be instead a
bounce trajectory from $\phi=0$ to $\phi_{1}$ and from $\phi_{2}$ to
$\pi$, where $\sin^{2}(\phi_{1,2})=(1-h^{2})/\lambda$. We then
consider the instanton given by the trajectory which consists of three
parts:
\begin{enumerate}
\item motion in imaginary time from $0$ to ${\phi}_{1}$,
\item motion in real time from ${\phi}_{1}$ to ${\phi}_{2}$,
\item motion in imaginary time from ${\phi}_{2}$ to ${\pi}$,
\end{enumerate}
and another instanton given by the trajectory going in the opposite
direction. It is clear from the shape of the potential that all three
parts of the trajectory join smoothly, because $\phi(\tau)$ and
$\dot{\phi}_{\tau}(\tau)$ coincide at the joints.  Note that the
real-time part of the trajectory still corresponds to the virtual
rotation of the magnetic moment, because for that part of the
trajectory $\cos(\theta)>1$, as can be seen from the classical
equations of motion that follow from the
Lagrangian~(\ref{lagrang}).

Using the energy conservation
\begin{equation}
  {\cal H}(\phi,\dot{\phi}_{\tau'}) = \frac{1}{2} M(\phi)
  \dot{\phi}^{2}_{\tau'} - V(\phi) = 0\;,
\end{equation}
one obtains
\begin{equation}
  \label{eq:eneconserv}
  \dot{\phi}^{2}_{\tau'} = \frac{2 V(\phi)}{M(\phi)} = (1-h^{2})
  \sin^{2}(\phi) \left[1 - \frac{\lambda}{1-h^{2}} \sin^{2}(\phi)
  \right]\;.
\end{equation}
This can be used to compute the second integral in eq.\ 
(\ref{effaction}).  For $h_{2}<h<1$, the real part of the
imaginary-time action is given by the contribution of the bounce
trajectories from $0$ to $\phi_{1}$ and from $\phi_{2}$ to $\pi$.
Integration gives
\begin{eqnarray}
  \label{realI}
  {\rm Re}\left(\frac{I}{\hbar}\right) &=& 2S\lambda^{1/2}
  \int_{[0,\phi_{1}]\cup[\phi_{2},\pi]} d\phi \left[2 M(\phi) V(\phi)
  \right]^{1/2} \nonumber \\
  &=& 4S\left[z-\left(\frac{1}{a^{2}}+1\right)^{1/2}
    \tanh^{-1}\left(\frac{a^{2}-b^{2}}{1+a^{2}}\right)^{1/2}\right]\;,
\end{eqnarray}
where
\[
  z = \cosh^{-1}\left(\frac{a}{b}\right)\;,\quad a^{2} =
  \frac{1-\lambda} {h^{2}-(1-\lambda)}\;,
\]
\begin{equation}
  b^{2}=\frac{1-\lambda}{\lambda} = \frac{k_{\perp}}{k_{\parallel}}\;.
\end{equation}
Equation\ (\ref{realI}) is the WKB exponent for the tunnel splitting.

In the uniaxial limit ($\lambda{\rightarrow}1$), one obtains
\begin{equation}
  \label{eq:uniaxial}
  {\rm Re}\left(\frac{I}{\hbar}\right) = 4S\left[\cosh^{-1}
    \left(\frac{1}{h}\right)- (1-h^{2})^{1/2}\right]\;,
\end{equation}
in full agreement with ref.\ \cite{garmarchu98}.  
This formula is correct in the entire field range $0<h<1$. In the
limit of small barrier\cite{chugun88}, $h=1-\epsilon$ with
${\epsilon}{\rightarrow}0$, it gives
\begin{equation}
  \label{eq:uni_eps}
  {\rm Re}\left(\frac{I}{\hbar}\right) =
  \frac{8\sqrt{2}}{3}S{\epsilon}^{3/2}\;.
\end{equation}

For $h<h_{2}$, the imaginary part of the action for the trajectory
going from $\phi=0$ to $\phi=\pi$ is given solely by the topological
term $(i/\hbar) S I_{WZB}^{\pm}$.  For $h_{2}<h<1$, however, we have
another contribution coming from the second integral in eq.\ 
(\ref{effaction}) evaluated for the real-time motion between
$\phi_{1}$ and $\phi_{2}$. Both contributions combine into
\begin{eqnarray}
  \label{eq:imI}
  {\rm Im}\left(\frac{I}{\hbar}\right) &=& \frac{i}{\hbar}S
  I_{WZB}^{\pm} \pm S\lambda^{1/2} \int_{\phi_{1}}^{\phi_{2}} d\phi
  \left[2 M(\phi) V(\phi) \right]^{1/2}
  \nonumber \\
  &=& \frac{i}{\hbar}S I_{WZB}^{\pm} \mp i\pi S \left(1 -
    \frac{h}{h_{2}}\right) = 0\;.
\end{eqnarray}
Remarkably, the real-time motion exactly cancels the contribution of
the Wess-Zumino-Berry phase.  Consequently, the freezing of tunneling
does not occur at high fields ($h_{2}<h<1$) and topological
oscillations are suppressed, as confirmed by our numerical
diagonalization of the Hamiltonian at $S=10$,
figs.~\ref{frzfig1}--\ref{frzfig3}. In each of the
figs.~\ref{frzfig1}--\ref{frzfig3} the field $h_{2}$ seem to coincide
with the inflexion point on the envelope curve.

Formally, one could argue that at $h_{2}<\,h<\,1$ the instantons
connecting the ${\phi}={\pm}{\pi}/2$ global minima of the effective
potential shown in fig.~\ref{frzfig4} should be used to compute the
tunneling splitting of the ground state. This would be conceptually
incorrect, however, because tunneling in the biaxial model occurs
between the energy minima ${\phi}=0,{\pi}$, which are the only
classical energy minima of the energy. Besides, the tunneling
splitting computed via the ${\phi}={\pm}{\pi}/2$ instantons is
different from the one given above and checked through the numerical
diagonalization of the Hamiltonian.

One should notice that for $h_{2}<h<1$ the particle spends finite real
time under the barrier, though the motion is still virtual because
$\cos(\theta)>1$. This time is given by
$T=2\pi/[\gamma[(1-h^{2})k_{\parallel} (k_{\parallel}+
k_{\perp})]^{1/2}]$. It tends to the period of small oscillations
around the minimum at $\phi=\pi/2$ when $h\rightarrow h_{2}$, and
diverges when the field approaches the critical value $h=1$. 
Notice also that complex instantons resembling our instantons have
been recently introduced in a rotating black hole
problem\cite{booman98}.

Taking the average of the measured values of the
parameters\cite{Barra,Caciuffo,Sangregorio}, we get $\lambda \approx
0.72$ for Fe$_8$, which gives $h_{2} \approx 0.53$, and $H_{c}\approx
5.1\un{T}$. Then the Garg's oscillations must exist at $H<H_{2}\approx
2.7\un{T}$, while above $2.7\un{T}$ the tunnel splitting must grow
monotonically with the field. This prediction can be tested in EPR and
inelastic neutron scattering experiments on well oriented crystals of
Fe$_{8}$.

\acknowledgments
The work of EMC has been supported by the U.S. National Science
Foundation under the Grant No.\ DMR-9024250.
The work of XMH has been supported by the Comissionat per a
Universitats i Recerca of the Generalitat de Catalunya.




\end{document}